\documentclass[12pt]{iopart}

\usepackage{iopams,citesort,graphicx,tabularx,setspace}  

\begin{document}

\newcommand{\rd}{\mathrm{d}}
\newcommand{\ri}{\mathrm{i}}
\newcommand{\p}{\partial}
\newcommand{\px}{{\bi{x}}}
\newcommand{\expct}[1]{\langle #1 \rangle}
\newcommand{\Expct}[1]{\left\langle #1 \right\rangle}
\newcommand{\cum}[1]{\langle #1 \rangle_\mathrm{c}}
\newcommand{\Cum}[1]{\left\langle #1 \right\rangle_\mathrm{c}}
\newcommand{\diff}[2]{\frac{\mathrm{d} #1}{\mathrm{d} #2}}
\newcommand{\prt}[2]{\frac{\partial #1}{\partial #2}}
\newcommand{\const}{\mathrm{const.}}
\renewcommand{\(}{\left(}
\renewcommand{\)}{\right)}
\renewcommand{\[}{\left[}
\renewcommand{\]}{\right]}
\newcommand{\im}{\mathrm{Im}}
\newcommand{\re}{\mathrm{Re}}
\newcommand{\unit}[1]{\,\mathrm{#1}}
\newcommand{\tmin}{t_\mathrm{min}}
\newcommand{\tmax}{t_\mathrm{max}}
\newcommand{\tod}{\stackrel{\mathrm{d}}{\to}}

\newcommand{\figref}[1]{figure~\ref{#1}}
\newcommand{\pref}[1]{(\ref{#1})}
\newcommand{\eqref}[1]{(\ref{#1})}

\title[$1/f^\alpha$ power spectrum in KPZ]{$1/f^\alpha$ power spectrum in the Kardar-Parisi-Zhang universality class}

\author{Kazumasa A Takeuchi}

\address{Department of Physics, Tokyo Institute of Technology,\\
 2-12-1 Ookayama, Meguro-ku, Tokyo 152-8551, Japan.}
\ead{kat@kaztake.org}

\begin{abstract}
The power spectrum of interface fluctuations in the $(1+1)$-dimensional
 Kardar-Parisi-Zhang (KPZ) universality class is studied
 both experimentally and numerically.
The $1/f^\alpha$-type spectrum is found and characterized
 through a set of ``critical exponents'' for the power spectrum.
The recently formulated ``aging Wiener-Khinchin theorem''
 accounts for the observed exponents.
Interestingly, the $1/f^\alpha$ spectrum in the KPZ class turns out to
 contain information on a universal distribution function
 characterizing the asymptotic state of the KPZ interfaces,
 namely the Baik-Rains universal variance.
It is indeed observed in the presented data,
 both experimental and numerical,
 and for both circular and flat interfaces,
 in the long time limit.
\end{abstract}


\section{Introduction}

The power spectrum%
\footnote{
In this paper, parameters of a function may be specified
 after its arguments, separated by a semicolon. 
These parameters may be dropped for simplicity.
}
\begin{equation}
 S(\omega; T) = \frac{1}{T}\left| \int_0^T X(t) \e^{-\ri\omega t} \rd t \right|^2  \label{eq:DefPowSpec}
\end{equation}
 is a useful tool to characterize fluctuating signals $X(t)$ in general.
Curiously, there have been reported a vast variety of systems showing
 a power law in $S(\omega)$ at low frequencies,
\begin{equation}
 S(\omega) \sim 1/\omega^\alpha  \qquad (0 < \alpha < 2) \label{eq:1/f}
\end{equation}
 ranging from solid-state physics (vacuum tubes, semiconductors, etc.) \cite{Dutta.Horn-RMP1981,Wong-MR2003,Musha.etal-book1992}
 to fluid mechanics (e.g., turbulence), life science (e.g., heart beats)
 and other branches of science and technology
 \cite{Mandelbrot-book1999,Musha.etal-book1992}.
While the generic term ``$1/f$ noise''
 or more generically ``$1/f^\alpha$ noise'' was coined to this phenomenon,
 various underlying mechanisms have been proposed in the literature.

Theoretical approaches to the $1/f^\alpha$ noise can be classified, roughly,
 into those based on \textit{stationary} processes and
 on non-stationary, or \textit{aging} ones.
Here the stationarity indicates that the correlation function,
 defined with the ensemble average
\begin{equation}
 C(\tau; t) = \expct{X(t+\tau)X(t)},  \label{eq:DefCorrFunc}
\end{equation}
 does not depend on $t$.
In this case the power spectrum $S(\omega; T)$
 is a function of $\omega$ solely, and becomes simply
 the Fourier transform of $C(\tau)$,
 thanks to the celebrated Wiener-Khinchin theorem \cite{Kubo.etal-Book1991}.
Since this simplifies analysis, and also because other kinds of noise such as
 Johnson-Nyquist noise and shot noise \cite{Kobayashi-PJAB2016}
 have been successfully described under the assumption of stationarity,
 the same assumption has also been often adopted in models of $1/f^\alpha$ noise
 for solid-state systems \cite{Dutta.Horn-RMP1981,Wong-MR2003}.
On the other side, recent experiments have shown
 examples of $1/f^\alpha$ noise in intermittent systems,
 such as blinking quantum dots \cite{Pelton.etal-APL2004,Pelton.etal-PNAS2007,Frantsuzov.etal-NL2013,Sadegh.etal-NJP2014} and nanoscale electrodes \cite{Krapf-PCCP2013}, which were shown to be non-stationary, or \textit{aging}.
The $1/f^\alpha$ noise in those systems was accounted for by models
 based on the intermittent dynamics \cite{Niemann.etal-PRL2013,Sadegh.etal-NJP2014,Krapf-PCCP2013},
 and the same line of analysis was also applied to
 turbulent fluid \cite{Herault.etal-EL2015,Herault.etal-JSP2015} and 
 fluctuating electroconvection of liquid crystal
 \cite{Silvestri.etal-PRL2009}.
More generally, since the power law \pref{eq:1/f}
 would naturally suggest scale invariance in time scales,
 $1/f^\alpha$ noise is widely expected in scale-invariant processes
 such as coarsening \cite{Bray-AP1994},
 self-organized criticality \cite{Bak.etal-PRL1987,Turcotte-RPP1999},
 and growth processes \cite{Barabasi.Stanley-Book1995},
 all of which are examples of aging systems.
The Wiener-Khinchin theorem cannot be used in aging systems,
 but recently an analogous relationship between the correlation function
 and the power spectrum was derived for aging systems,
 named the ``aging Wiener-Khinchin theorem''
 \cite{Leibovich.Barkai-PRL2015,Dechant.Lutz-PRL2015,Leibovich.etal-PRE2016}.

In this work, taking an example of scale-invariant fluctuations
 of growing interfaces,
 we characterize the power spectrum by a set of recently proposed
 ``critical exponents'' \cite{Sadegh.etal-NJP2014,Leibovich.Barkai-PRL2015,Dechant.Lutz-PRL2015,Leibovich.etal-PRE2016}.
This analysis has an advantage that
 one does not make any \textit{a priori} assumption about the stationarity;
 instead, with the obtained exponents, one can judge
 whether the system is stationary or not, and,
 if aging is there, determine its time dependence.
For the systems we study here, the observed exponents indeed indicate
 the relevance of aging, and they are successfully accounted for
 by the aging Wiener-Khinchin theorem.
Moreover, this theorem turns out to unveil
 certain universal fluctuation property of the studied systems,
 namely the universal variance of the Baik-Rains distribution
 \cite{Baik.Rains-JSP2000} for the Kardar-Parizi-Zhang (KPZ)
 universality class \cite{Kardar.etal-PRL1986,Barabasi.Stanley-Book1995,Kriecherbauer.Krug-JPA2010,Corwin-RMTA2012,HalpinHealy.Takeuchi-JSP2015}.
This demonstrates that, at least in the problem of growing interfaces,
 the aging Wiener-Khinchin theorem can be used as a practical tool
 to characterize the systems of interest.

The paper is organized as follows.
Section~\ref{sec:KPZ} is devoted to a brief introduction
 of growing interface fluctuations and the KPZ class,
 with some emphasis on remarkable theoretical developments
 attained for the $(1+1)$-dimensional case.
The systems to study, either experimentally or numerically,
 are described in section~\ref{sec:systems}.
The results on the power spectrum, characterized
 through the critical exponents and the aging Wiener-Khinchin theorem,
 are presented in section~\ref{sec:results}.
Concluding remarks are given in section~\ref{sec:conclusion}.

\section{Growing interfaces and KPZ}  \label{sec:KPZ}

\begin{figure}[t]
 \centering
 \includegraphics[width=\hsize,clip]{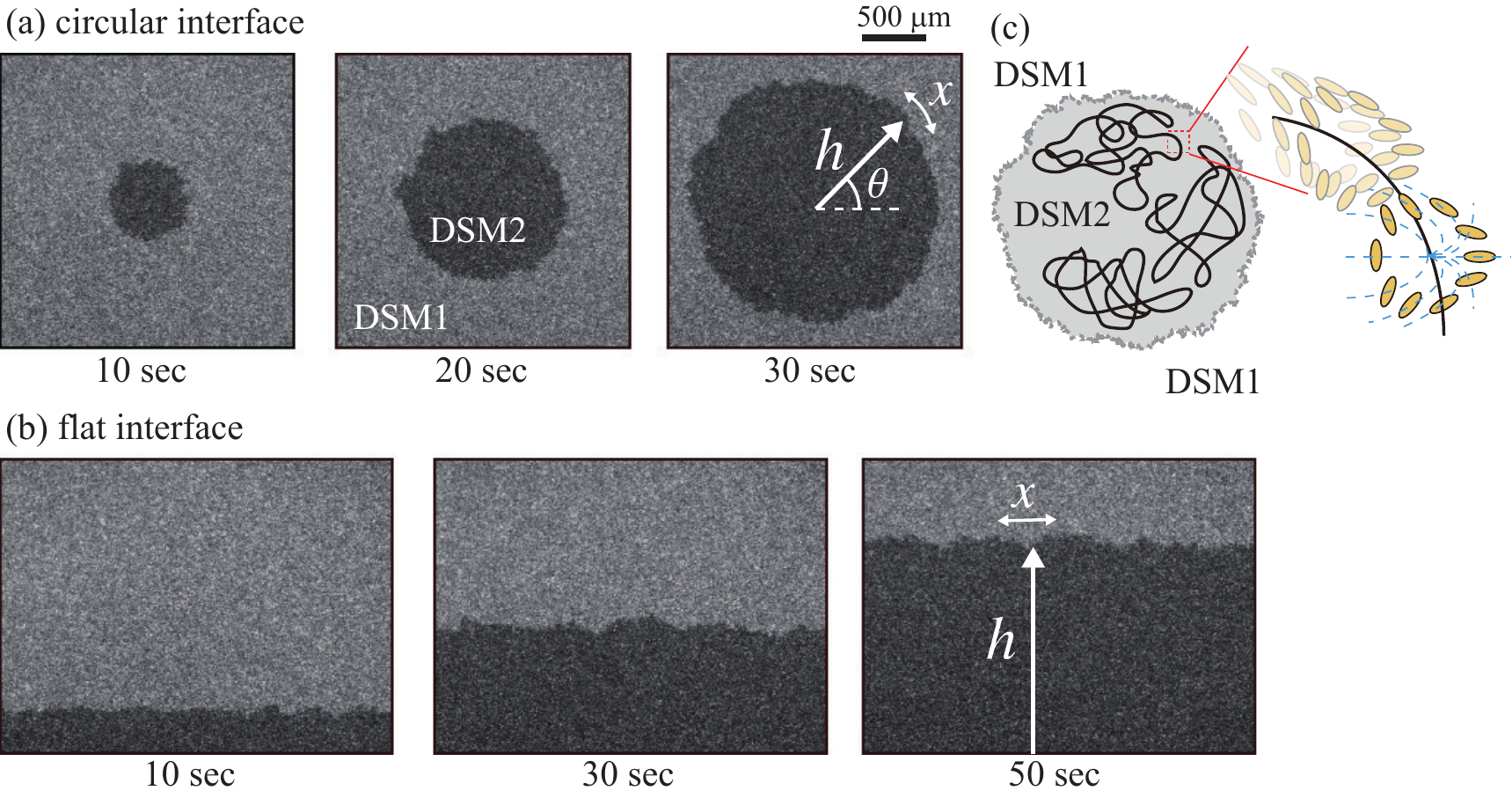}
 \caption{
Growing interfaces in liquid-crystal turbulence \cite{Takeuchi.Sano-PRL2010,Takeuchi.etal-SR2011,Takeuchi.Sano-JSP2012}, separating an expanding DSM2 region (black) and the surrounding DSM1 region (gray). See section~\ref{sec:LCconvection} for descriptions of the experimental system and the DSM1/DSM2 turbulence. Starting from a point DSM2 nucleus, one obtains a circular interface (a), while from a line a flat interface is generated (b). The time is measured from the emission of laser pulses used to trigger DSM2. See also Supplementary Movies of reference \cite{Takeuchi.etal-SR2011}. (c) Sketch of the DSM1 and DSM2 states. DSM2 consists of a large density of line defects, specifically disclinations, in the liquid-crystal director field. The panels (a) and (b) were reprinted with adaptation from Fig.~3(a) of reference \cite{Takeuchi.Sano-JSP2012} and the panel (c) from Fig.~1(b) of the same reference, with permission of Springer.
}
 \label{fig1}
\end{figure}%

When a system consists of two regions bordered
 by a well-defined interface, and if one of the two regions expands
 in a fluctuating manner, the interface typically develops intricate
 winding structure (see, e.g., \figref{fig1}ab).
Examples abound from physics to biology and chemistry,
 as well as from nanoscales to macroscales, including
 surfaces of deposited solid films
 \cite{Palasantzas.etal-SS2002,Almeida.etal-PRB2014},
 growing clusters of liquid-crystal turbulence
 \cite{Takeuchi.Sano-PRL2010,Takeuchi.etal-SR2011,Takeuchi.Sano-JSP2012},
 fronts of smoldering papers
 \cite{Maunuksela.etal-PRL1997,Myllys.etal-PRE2001},
 expanding colony of bacteria \cite{Wakita.etal-JPSJ1997},
 propagation of chemical waves in disordered flow \cite{Atis.etal-PRL2015},
 etc. (see also \cite{Barabasi.Stanley-Book1995,Takeuchi-JSM2014}).
Interestingly, the observed patterns are generically scale-invariant,
 without fine tuning of experimental conditions and/or parameters
 \cite{Barabasi.Stanley-Book1995}: in terms of the local height of interface,
 $h(x,t)$, as a function of lateral coordinates $x$ and time $t$
 (\figref{fig1}ab)%
\footnote{
Note that, for the circular interfaces (e.g., \figref{fig1}a),
 the lateral coordinate is more appropriately specified
 by the azimuth $\theta$, which remains constant
 along the mean growth direction.
With this, the coordinate $x$ (of dimension of length) can be formally defined
 by $x = R(t)\theta$, using mean radius $R(t)$ at time $t$.
},
 its fluctuation $\delta h(x,t) \equiv h(x,t) - \expct{h(x,t)}$ is
 statistically invariant under certain set of coordinate rescaling,
 $t \to bt$, $x \to b^{1/z} x$, $\delta h \to b^{\beta} \delta h$.
This is analogous to dynamic critical behavior,
 with $\beta$ and $z$ playing the role of the critical exponents
 and $h(x,t)$ being the order parameter.
As a result, such growing interfaces may be described
 by a set of universal scaling laws,
 with the critical exponents and scaling functions
 specific to each universality class.
The KPZ class is one such universality class,
 known to describe the simplest generic case \cite{Barabasi.Stanley-Book1995},
 with a variety of examples from theoretical models and experiments; in fact,
 the above-mentioned experiments were all shown to be in the KPZ class,
 at least in some parameter space
 and with the materials chosen in the studies%
\footnote{
There also exist many experimental examples of growing interfaces
 that are \textit{not} in the KPZ class
 \cite{Barabasi.Stanley-Book1995,Takeuchi-JSM2014}.
Although some of those systems have macroscopic ingredients
 that can generally affect the universality class, such as quenched disorder,
 it remains challenging to predict whether a given experimental system
 is in the KPZ class or not.
}.

Recently, the KPZ class became crucial in the studies of non-equilibrium
 scaling laws, when it turned out to be analytically tractable in many aspects
 for the $(1+1)$-dimensional case \cite{Kriecherbauer.Krug-JPA2010,Corwin-RMTA2012,HalpinHealy.Takeuchi-JSP2015}, 
 i.e., for one-dimensional interfaces growing in two-dimensional space.
In this case, the scaling exponents are known to be
 $z = 3/2$ and $\beta = 1/3$
 \cite{Kardar.etal-PRL1986,Barabasi.Stanley-Book1995};
 the fluctuation amplitude increases as $\delta h \sim t^{1/3}$,
 with the lateral correlation length $\xi \sim t^{2/3}$.
The height $h(x,t)$ grows, therefore, as
\begin{equation}
 h(x,t) \simeq v_\infty t + (\Gamma t)^{1/3} \chi_{x,t},
  \label{eq:Rescaling}
\end{equation}
 with constant parameters $v_\infty, \Gamma$
 and a rescaled random variable $\chi_{x,t}$,
 which carries all relevant information of KPZ interface fluctuations.
The recent theoretical developments then provided exact solutions
 for the one-point distribution of $\chi_{x,t}$,
 as well as its spatial correlation,
 revealing surprising link to random matrix theory, combinatorics,
 and quantum integrability \cite{Prahofer.Spohn-PRL2000,Kriecherbauer.Krug-JPA2010,Corwin-RMTA2012,HalpinHealy.Takeuchi-JSP2015}.
Remarkably, the results turned out to be classified into a few
 \textit{universality subclasses} \cite{Prahofer.Spohn-PRL2000},
 determined by the global geometry of interfaces,
 or equivalently by the initial condition (table \ref{tbl:subclass}).
For circular interfaces growing from a point nucleus (\figref{fig1}a and \figref{fig2}b),
 the asymptotic distribution is given by the Tracy-Widom distribution
 for the Gaussian unitary ensemble (GUE), originally introduced to describe
 fluctuations of the largest eigenvalue of GUE random matrices
 \cite{Tracy.Widom-CMP1994};
in other words, $\chi_{x,t} \tod \chi_2$,
 where ``$\tod$'' denotes convergence in the distribution
 and $\chi_2$ is the standard random variable
 for the GUE Tracy-Widom distribution.
This case is called the circular subclass,
 a subset of the KPZ class for circular interfaces.
Similarly, the flat subclass describes globally flat interfaces
 starting from a straight line (\figref{fig1}b).
It is characterized by the Tracy-Widom distribution
 for the Gaussian orthogonal ensemble (GOE) \cite{Tracy.Widom-CMP1996},
 or more precisely, $\chi_{x,t} \tod \chi_1$
 with $\chi_1$ being the GOE Tracy-Widom variable multiplied by $2^{-2/3}$
 \cite{Prahofer.Spohn-PRL2000}.
Another established subclass is associated with the Brownian initial condition,
 i.e., $h(x,0) = B_x$ with Brownian motion $B_t$,
 which is known to coincide with the asymptotic profile of interfaces
 in the $(1+1)$-dimensional KPZ class \cite{Barabasi.Stanley-Book1995}%
\footnote{
Therefore, this subclass is often called the ``stationary subclass''
 in the literature, while here we call it the Brownian subclass,
 in order to avoid confusion with stationary processes. 
The diffusion coefficient of the Brownian motion $B_t$
 is chosen in such a way that
 it matches with that of the asymptotic interface profile.}.
This subclass is then characterized by another distribution
 called the Baik-Rains distribution \cite{Baik.Rains-JSP2000}
 ($\chi_{x,t} \tod \chi_0$).
The spatial correlation was also solved,
 and given by time correlation of certain stochastic process,
 called the Airy$_2$, Airy$_1$, and Airy$_\mathrm{stat}$ process
 for the circular, flat, and Brownian subclass, respectively
  \cite{Kriecherbauer.Krug-JPA2010,Corwin-RMTA2012,HalpinHealy.Takeuchi-JSP2015}.
While other subclasses may also exist, 
 these three are the main subclasses
 constituting the $(1+1)$-dimensional KPZ class,
 characterized by the same critical exponents,
 yet by the different non-Gaussian distributions and correlation functions.

\begin{table}[t]
 \begin{center}
  \caption{Three representative subclasses for the $(1+1)$-dimensional KPZ class  \cite{Kriecherbauer.Krug-JPA2010,Corwin-RMTA2012,HalpinHealy.Takeuchi-JSP2015}.}
  \label{tbl:subclass}
  \catcode`?=\active \def?{\phantom{0}}
  {\small
  \begin{tabular}{l|ccc} \hline
  KPZ subclass & circular & flat & Brownian \\ \hline \\[-12pt]
  \shortstack{initial condition$^{\rm a}$ \\ (typical) \vspace{-14pt}} & $h(x,0) \!=\! \left\{ \!\!\!\begin{array}{ll} 0 & \!(x=0) \\ -\infty & \!(x \neq 0) \end{array} \right.\!\!\!$ & $h(x,0)=0$ & $h(x,0) = B_x$ \\
  & (or point nucleus) \\ [4pt]
  exponents & \multicolumn{3}{c}{$z = 3/2$ and $\beta = 1/3$ for all subclasses} \\[4pt]
  distribution & GUE Tracy-Widom ($\chi_2$) & GOE Tracy-Widom ($\chi_1$) & Baik-Rains ($\chi_0$) \\
 \footnotesize (mean, var, s, k)$^{\rm b}$ & \footnotesize (-1.77, 0.813, 0.224, 0.093) & \footnotesize (-0.760, 0.638, 0.293, 0.165)$^{\rm c}$ & \footnotesize (0, 1.15, 0.359, 0.289) \\[4pt]
  spatial process & Airy$_2$ process & Airy$_1$ process & Airy$_\mathrm{stat}$ process \\ \hline
  \end{tabular}
  }
 \begin{spacing}{0.8}
 \raggedright\footnotesize $^{\rm a}$ The listed initial conditions are those typically used in the literature. Other conditions that share the same global symmetry are expected to lead to the same subclass.\\
 \raggedright\footnotesize $^{\rm b}$ Mean $\expct{\chi_i}$, variance $\cum{\chi_i^2}$, skewness $\cum{\chi_i^3}/\cum{\chi_i^2}^{3/2}$ and kurtosis $\cum{\chi_i^4}/\cum{\chi_i^2}^{2}$ are shown, where $\cum{\chi_i^n}$ denotes the $n$th-order cumulant of $\chi_i$. These are cited from \cite{Prahofer.Spohn-PRL2000}, in which more precise values are given.\\
 \raggedright\footnotesize $^{\rm c}$ As noted in the main text, $\chi_1$ differs from the usual definition of the GOE Tracy-Widom variable by the factor $2^{-2/3}$. The values of the cumulants change accordingly.
 \end{spacing}
 \end{center}
\end{table}

In contrast to these remarkable developments
 on the distribution and spatial correlation functions \cite{Kriecherbauer.Krug-JPA2010,Corwin-RMTA2012,HalpinHealy.Takeuchi-JSP2015},
 much less is known theoretically about the time correlation;
 analytical treatment of two-time quantities
 was made only very recently \cite{Ferrari.Spohn-SIG2016,Johansson-CMP2017,DeNardis.etal-PRL2017} and many other aspects of time correlation,
 in particular persistence properties,
 observed experimentally \cite{Takeuchi.Sano-JSP2012,Takeuchi.Akimoto-JSP2016}
 and numerically \cite{Kallabis.Krug-EL1999,Singha-JSM2005,Takeuchi-JSM2012,Takeuchi.Akimoto-JSP2016} remain to be explained.
Here we aim to characterize how the local fluctuation of interface
 $\delta h(x,t)$ evolves in time,
 using the power spectrum and related analysis developed
 in the context of the $1/f^\alpha$ noise.
Note that, while KPZ dynamics is clearly non-stationary, or aging,
 depending explicitly on time $t$
 measured from the start of the growth process,
 this non-stationarity will not be assumed \textit{a priori}
 in the data analysis.
Concerning the systems to study,
 we use the experiment on liquid-crystal turbulence
 \cite{Takeuchi.Sano-PRL2010,Takeuchi.etal-SR2011,Takeuchi.Sano-JSP2012}
 (\figref{fig1}; circular and flat cases),
 as well as an off-lattice version of the Eden model \cite{Takeuchi-JSM2012}
 (\figref{fig2}; circular case)
 and a discrete version of the polynuclear growth (PNG) model
 \cite{Sasamoto.Imamura-JSP2004} (flat case),
 described in the following section.
These models are not meant to describe the liquid-crystal turbulence,
 but studied instead to test universality of the results within the KPZ class.

\section{Systems}  \label{sec:systems}

\subsection{Liquid-crystal turbulence}  \label{sec:LCconvection}

Nematic liquid crystal
 with negative dielectric anisotropy $\epsilon_\parallel < \epsilon_\perp$
 and positive conductivity anisotropy $\sigma_\parallel > \sigma_\perp$
 is known to develop convection,
 when confined between two plates
 and subjected to an alternating electric field
 with a relatively low frequency, 
 driven by the Carr-Helfrich effect \cite{deGennes.Prost-Book1995}.
Similarly to the thermal convection, this electroconvection undergoes
 a series of transitions in the convection pattern as the driving strength
 -- the amplitude of the voltage applied to the system -- increases.
Eventually, the system reaches regimes of turbulence,
 or more precisely, spatiotemporal chaos,
 called the dynamic scattering modes (DSM) 1 and 2.
They are distinct in the density of topological defects,
 specifically, disclinations, in the liquid-crystal director field
 (\figref{fig1}c):
 while DSM1 has no sustained disclinations, DSM2 consists of a large density
 of densely entangled disclinations.
Therefore, DSM2 scatters light more strongly, hence
 looks darker in the transmitted light images (see \figref{fig1}).
Under sufficiently high applied voltage, these disclinations are kept dense,
 stretched, and randomly transported
 by local turbulent flow of the electroconvection. 
This leads in macroscopic scales to fluctuating growth of the DSM2 region,
 which takes over the metastable DSM1 state (\figref{fig1}).

In our setup, detailed in the author's past publication
 \cite{Takeuchi.Sano-JSP2012}, the initial DSM2 region was generated
 by shooting ultraviolet laser pulses to the sample.
If pulses are focused on a point, a DSM2 point nucleus is created,
 which then grows, forming a circular interface (\figref{fig1}a).
If the laser beam profile is expanded along a line,
 longer than the width of the camera view field,
 DSM2 is generated on that line, producing a flat interface (\figref{fig1}b).
This allows us to study both circular and flat interfaces
 under the practically same experimental conditions,
 each case repeated nearly one thousand times
 to achieve high statistical accuracy.
For the applied voltage 26$\unit{V}$ used here,
 the scaling coefficients were estimated to be
 $v_\infty = 33.24(4) \unit{\mu{}m/s}$ and
 $\Gamma = 2.29(3) \times 10^3 \unit{\mu{}m^3/s}$ for the circular case,
 and  $v_\infty = 32.75(3) \unit{\mu{}m/s}$ and
 $\Gamma = 2.15(10) \times 10^3 \unit{\mu{}m^3/s}$ for the flat case
 \cite{Takeuchi.Sano-JSP2012},
 where the numbers in the parentheses
 indicate uncertainties in the last digit(s) of the estimates%
\footnote{
Although the parameters $v_\infty$ and $\Gamma$
 are expected to be independent of the interface geometry,
 the experimentally obtained estimates were slightly different. 
This is presumably because of small uncontrolled shifts 
 in the experimental conditions,
 during the period (8 days) between the two sets of the experiments.
}.
The circular and flat interfaces were found to show the hallmarks
 of the circular and flat KPZ subclasses, respectively,
 indicated in table \ref{tbl:subclass}.
In the following, the applied voltage is fixed at 26$\unit{V}$,
 but as long as the voltage is within a reasonable range
 for studying growing DSM2 interfaces, we can expect the same results
 (under proper rescaling) at different voltages \cite{Takeuchi.Sano-JSP2012}.

To evaluate the power spectrum \pref{eq:DefPowSpec},
 we use time series of $h(x,t)$ along the growth direction,
 i.e., with fixed $\theta$ for the circular case
 and fixed $x$ for the flat case (see \figref{fig1}ab).
Because the interfaces were not faithfully detected
 for the first few seconds from the laser emission,
 the time series were recorded for $\tmin \leq t \leq \tmax$,
 with $(\tmin, \tmax) = (2.0\unit{s}, 30.5\unit{s})$ for the circular case and 
 $(\tmin, \tmax) = (3.0\unit{s}, 63.0\unit{s})$ for the flat case.
Concerning the time resolution, 
 all images taken during the above time periods
 were analyzed in the present study,
 so that the time interval $\Delta t$ is $0.10\unit{s}$ for the circular case
 and $0.12\unit{s}$ for the flat case,
 improving the resolution used in the past work \cite{Takeuchi.Sano-JSP2012}.

\begin{figure}[t]
 \centering
 \includegraphics[width=\hsize,clip]{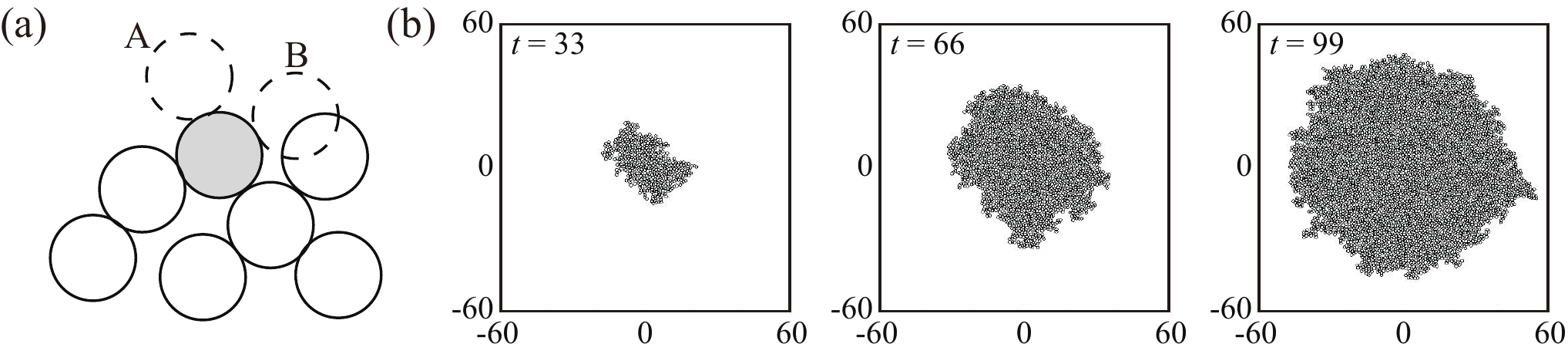}
 \caption{
Off-lattice Eden model. 
(a) Evolution rule. 
Suppose the gray particle is chosen from $N_\mathrm{part}(t)$ existing ones.
Then we attempt to put a new particle, in an angular direction randomly chosen
 from $[0, 2\pi)$. If it does not overlap with other particles
 (such as the position A in the figure), the attempt is adopted,
 otherwise (such as the position B) it is withdrawn.
(b) Typical evolution of an Eden cluster.
}
 \label{fig2}
\end{figure}%

\subsection{Off-lattice Eden model}

To corroborate the experimental results obtained in the present work,
 we also investigate simple models of growing interfaces,
 which are known to be in the KPZ class.
For the circular case, to attain better statistical accuracy,
 one needs an isotropically growing system
 so that the interface height $h(x,t)$ at all angular positions
 can be treated equally.
For this reason, here we use the off-lattice Eden model
 introduced in \cite{Takeuchi-JSM2012}.
This model deals with a cluster made of round particles of unit diameter,
 placed one by one in a two-dimensional space (\figref{fig2}a).
Suppose there are $N_\mathrm{part}(t)$ particles at time $t$.
Then, at each time step, one chooses a particle randomly from them,
 and attempts to introduce a new particle next to it,
 in a direction chosen randomly in the two-dimensional plane.
The new particle is placed if it does not overlap with any existing particle,
 otherwise the attempt is withdrawn.
In any case, time $t$ is increased by $1/N_\mathrm{part}(t)$.
Therefore, starting from a particle at the origin at $t=0$,
 one obtains a growing cluster, bordered by an on average circular interface
 (\figref{fig2}b).

To speed up simulations, particles without empty adjacent space
 were excluded from the count of $N_\mathrm{part}(t)$ (but still exist).
Similarly, since we are only interested in the interface,
 particles left inside the outermost perimeter were also regularly excluded
 (see \cite{Takeuchi-JSM2012} for details).
Time series were obtained from 100 independent realizations,
 in the range $t \leq \tmax = 50000$ with resolution $\Delta t = 1$.
For the scaling coefficients, we use $v_\infty = 0.51371$ and $\Gamma = 1.00$,
 taking the estimates from extensive simulations reported
 in \cite{Alves.etal-JSM2013}.

\subsection{Discrete PNG model}

For the flat case, since we do not need isotropic growth,
 a lattice model is more convenient and fast to simulate.
Here we use a discrete version of the PNG model
 with checkerboard updating, defined as follows.
In this model, the height $h(x,t)$ is integer,
 and both $x$ and $t$ are discretized
 with step size $\delta x$ and $\delta t$, respectively,
 i.e., $x = m \delta x$ and $t = n\delta t$ with integer $m$ and $n$.
Starting from the flat initial condition $h(x,0) = 0$,
 the height is updated by
\begin{equation}
 h(x, t+\delta t) = \max\{ h(x-\delta x, t), h(x,t), h(x+\delta x,t)\} + \eta(x,t),
\end{equation}
 where $\eta(x,t)$ is a random integer, independently drawn
 from the geometric distribution:
\begin{equation}
 \mathrm{Prob}[\eta(x,t) = k] = (1-p) p^k  \qquad (k = 0, 1, 2, \cdots)
\end{equation}
The update is made only for odd sites at odd times
 and for even sites at even times, i.e., $\mathrm{mod}(|n-m|,2) = 0$.
The advantage of this checkerboard updating is that
 the scaling coefficients are known exactly \cite{Sasamoto.Imamura-JSP2004},
 specifically,
\begin{equation}
 v_\infty = \frac{\sqrt{p}}{(1-\sqrt{p})\delta t}, \qquad
 \Gamma = \frac{\sqrt{p}(1+\sqrt{p})}{2(1-\sqrt{p})^3 \delta t}.
\end{equation}
By $\delta x = \delta t \to 0$ and $p = 2\rho \delta x \delta t$,
 one obtains the standard PNG model defined in continuous space and time,
 with nucleation rate $\rho$ and step speed 1.

In the present work, we use $\delta x = \delta t = 0.1$ and $\rho = 2$.
The lattice has $5 \times 10^5$ sites with the periodic boundary condition.
Time series were recorded only at even time steps,
 hence the resolution is $\Delta t = 2\delta t = 0.2$.
Simulations were carried out up to $\tmax = 10^6 \delta t$
 and 100 independent realizations were obtained.

\section{Results}  \label{sec:results}

\subsection{power spectrum and critical exponents}

\begin{figure}[t]
 \centering
 \includegraphics[width=\hsize,clip]{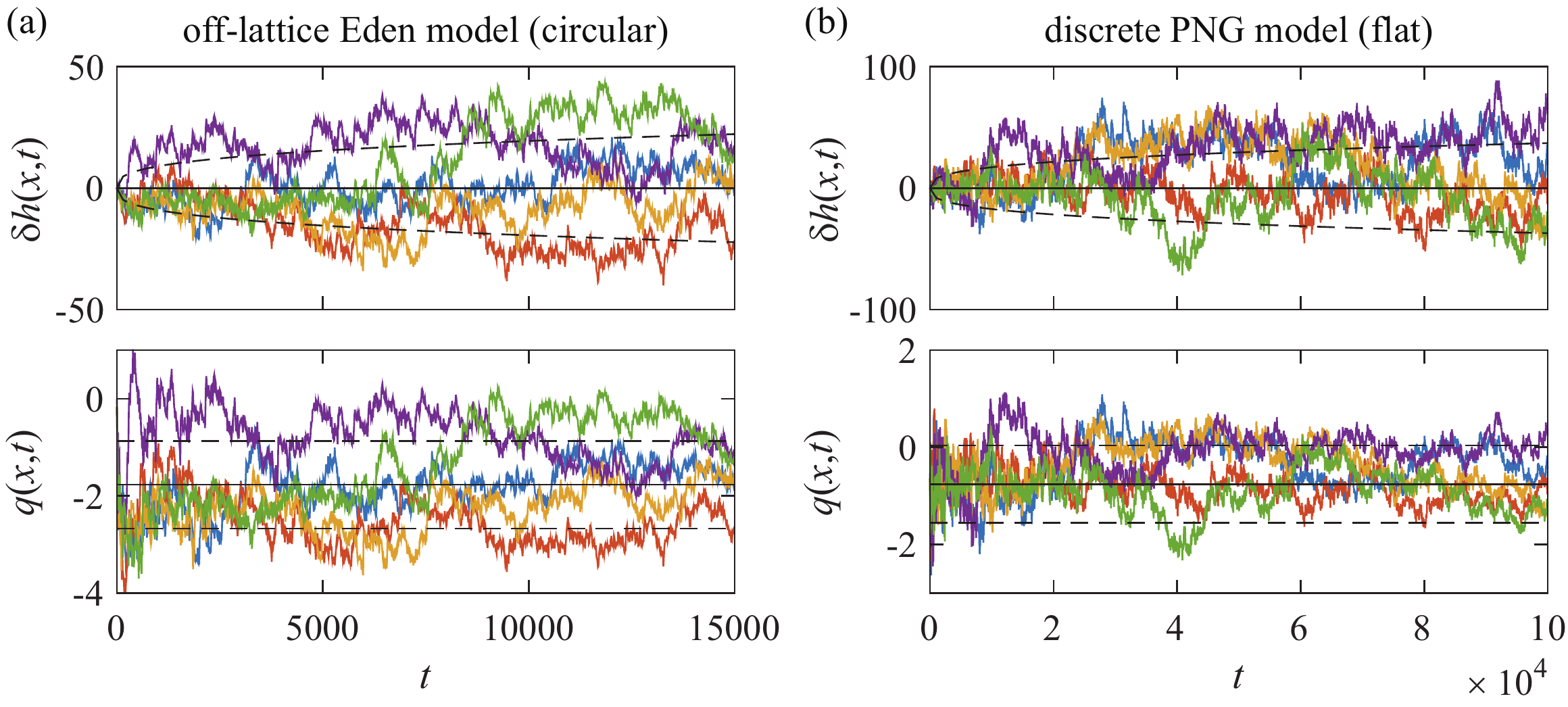}
 \caption{
Typical time series of $\delta h(x,t)$ and $q(x,t)$ for the circular (a) and flat (b) cases. Numerical data are shown here, for which much longer time series are available. The solid and dashed black lines indicate the mean value and the range of the standard deviation, respectively, valid for large $t$. It is given by $\pm (\Gamma t)^{1/3}\sqrt{\cum{\chi_i^2}}$ for $\delta h(x,t)$ and $\expct{\chi_i} \pm \sqrt{\cum{\chi_i^2}}$ for $q(x,t)$, with $i=2$ for the circular case and $i=1$ for the flat case.
}
 \label{fig3}
\end{figure}

In the present work, we shall consider time series of two different quantities,
 both reflecting the local fluctuation of the height $h(x,t)$
 at a fixed position
 (fixed $x$ for the flat case and fixed $\theta$ for the circular case;
 see \figref{fig1}ab).
The first quantity is the height fluctuation from the ensemble average%
\footnote{
Throughout section~\ref{sec:results},
 the ensemble average $\expct{\cdots}$ is obtained
 by averaging over realizations and spatial positions.},
 $\delta h(x,t) \equiv h(x,t) - \expct{h(x,t)}$, which grows as $t^{1/3}$
 (\figref{fig3}, top panels).
The second quantity is the rescaled height, defined by
\begin{equation}
 q(x,t) \equiv \frac{h(x,t) - v_\infty t}{(\Gamma t)^{1/3}} \simeq \chi_{x,t},
\end{equation}
 which remains $\mathcal{O}(1)$ (\figref{fig3}, bottom panels)
 with a well-defined asymptotic distribution,
 i.e., $q \tod \chi_i$
 with $i=1$ or 2 (see table \ref{tbl:subclass}).
Consequently, we study the following two power spectra,
 averaged over $x$ and realizations:
\begin{equation}
 S_h(\omega; T)
 = \Expct{\frac{1}{N \Delta t}\left| \sum_{n = 1}^N \delta h(x,t_n) \e^{-\ri\omega t_n} \Delta t \right|^2} \label{eq:DefPowSpecH}
\end{equation}
 and
\begin{equation}
 S_q(\omega; T)
 = \Expct{\frac{1}{N \Delta t}\left| \sum_{n = 1}^N q(x,t_n) \e^{-\ri\omega t_n} \Delta t \right|^2} \label{eq:DefPowSpecQ}
\end{equation}
 with $t_n = \tmin + (n-1)\Delta t$ ($\tmin = \Delta t$ for the simulations)%
\footnote{
Since aging processes do not have time translation symmetry,
 power spectrum in such processes depends not only on the measurement time $T$
 but also on the time to wait before the measurement, $\tmin$
 \cite{Bouchaud.etal-Inbook1997,Niemann.etal-PRL2013}.
Here we shall not consider this $\tmin$-dependence,
 because $\tmin$ is fixed, taken as small as possible,
 and $\tmin \ll T$ is satisfied.
Theoretically, the aging Wiener-Khinchin theorem that we shall use
 in this paper can be generalized
 for the case where $\tmin$ is not negligible
 \cite{Leibovich.Barkai-PC}.
},
 $t_N = T$, and $\omega$ being a multiple of $2\pi/N \Delta t$.

\begin{figure}[p]
 \centering
 \includegraphics[width=\hsize,clip]{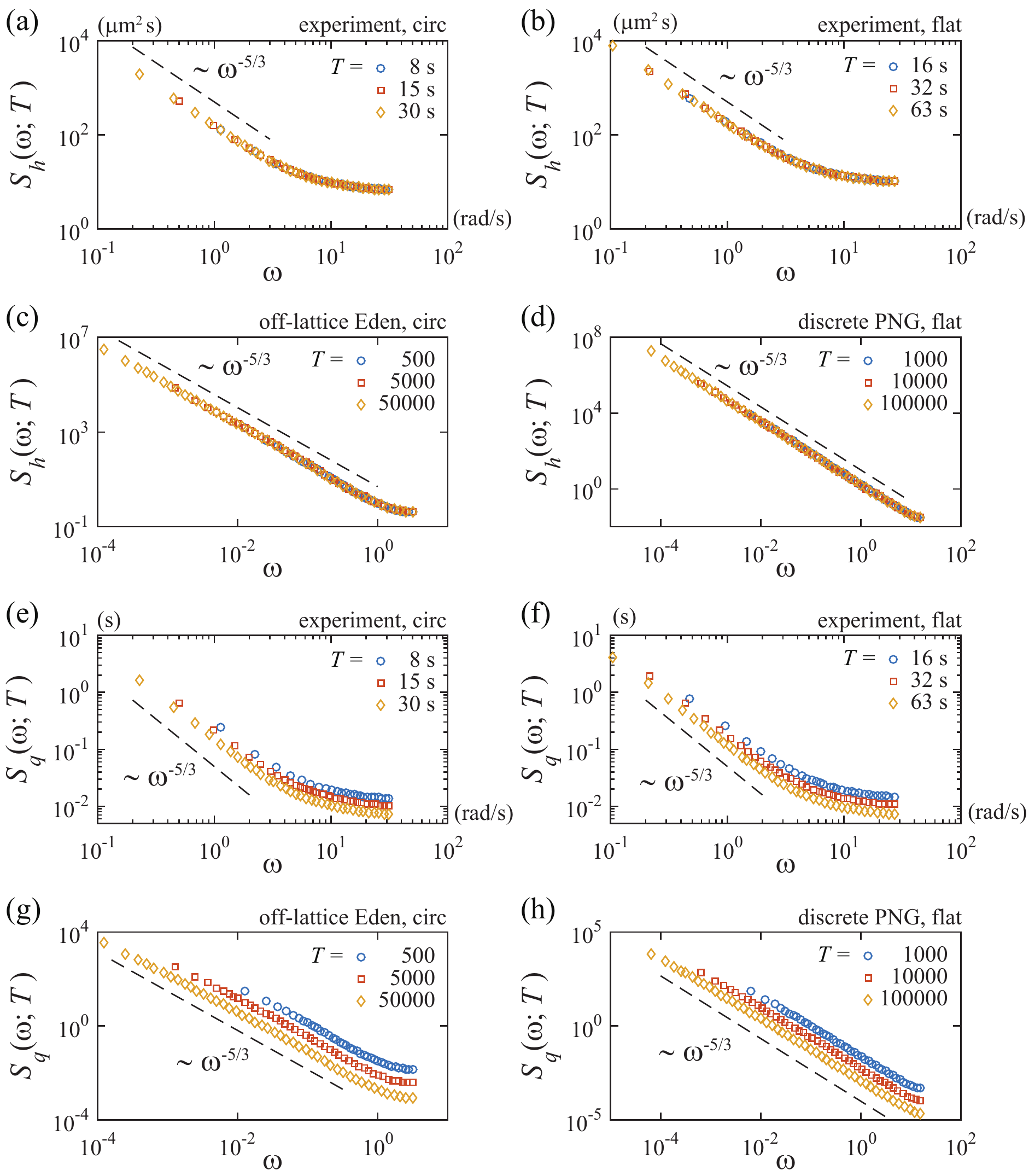}
 \caption{
Power spectra $S_h(\omega; T)$ and $S_q(\omega; T)$ for the liquid-crystal experiments [circular (a,e), flat (b,f)], the off-lattice Eden model [circular (c,g)], and the discrete PNG model [flat (d,h)]. The dashed lines are guides for the eyes showing the power law $S_h(\omega; T) \sim \omega^{-5/3}$, $S_q(\omega; T) \sim \omega^{-5/3}$.
}
 \label{fig4}
\end{figure}%

\begin{figure}[p]
 \centering
 \includegraphics[width=\hsize,clip]{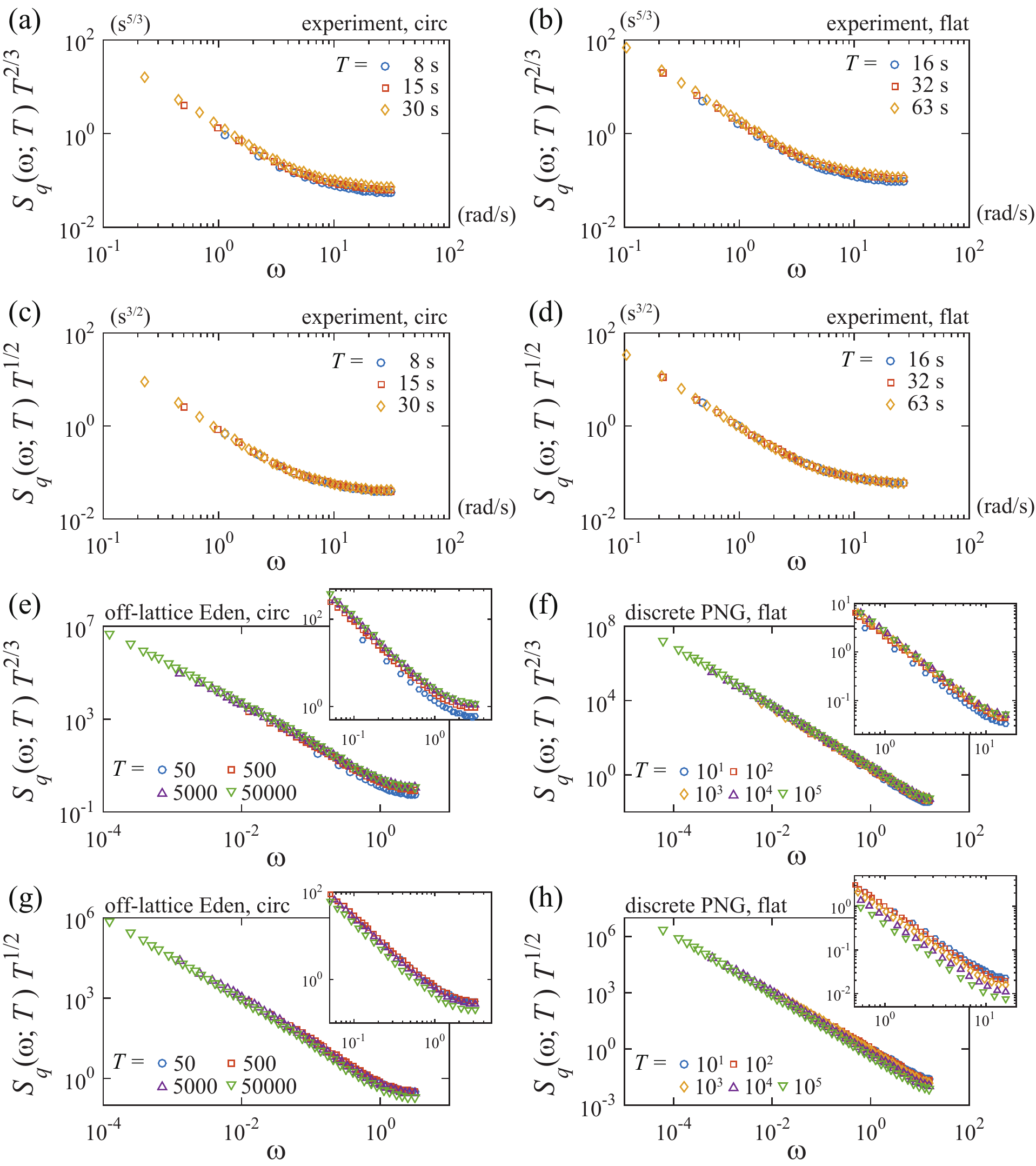}
 \caption{
Evaluation of the aging exponent $z_q$ for the power spectrum of the rescaled height, $S_q(\omega; T)$, for the liquid-crystal experiments [circular (a,c), flat (b,d)], the off-lattice Eden model [circular (e,g)], and the discrete PNG model [flat (f,h)]. The ordinates are $S_q(\omega; T) T^{z_q}$ with $z_q = 2/3$ (a,b,e,f) and $z_q = 1/2$ (c,d,g,h). For panels (e-h), the data are zoomed in and shown in the insets. Note that the data for large $T$ (upright and inverted triangles) overlap with $z_q = 2/3$ (insets of e,f), but not with $z_q = 1/2$ (insets of g,h). See also table~\ref{tbl:ExpRes}.
}
 \label{fig5}
\end{figure}%

Figure~\ref{fig4} shows the power spectra
 $S_h(\omega; T)$ (panel a-d) and $S_q(\omega; T)$ (panel e-h)
 with different $T$,
 for both the experiments (a,b,e,f) and the simulations (c,d,g,h)
 and for both the circular case (a,c,e,g) and the flat case (b,d,f,h).
In all cases, the power spectrum shows a power law at low frequencies,
 $S_h(\omega; T) \sim \omega^{-\alpha_h}$
 and $S_q(\omega; T) \sim \omega^{-\alpha_q}$ with $\alpha_h = \alpha_q = 5/3$,
 hence the $1/f^\alpha$-type spectrum is identified
 (see table~\ref{tbl:ExpRes} for the estimated exponent values).
Moreover, while $S_h(\omega; T)$ does not depend
 on the measurement time $T$ (\figref{fig4}a-d),
 $S_q(\omega; T)$ is found to decrease with increasing $T$ (\figref{fig4}e-h).
This may look somewhat counterintuitive
 if we recall $\delta h(x,t) \sim t^{1/3}$ and $q(x,t) \sim \mathcal{O}(1)$,
 but it can be clearly accounted for, as explained below.

\begin{table}[b]
 \begin{center}
  \caption{Experimental and numerical estimates of the critical exponents$^{\rm a}$.}
  \label{tbl:ExpRes}
  \catcode`?=\active \def?{\phantom{0}}
  {\footnotesize
  \begin{tabular}{l|cccc|cccc} \hline
   system$^{\rm b}$ & $\alpha_h$ & $z_h$ & $\mu_h$ & $\delta_h$ & $\alpha_q$ & $z_q$ & $\mu_q$ & $\delta_q$ \\ \hline
   exp. (circular) & 1.69(3) & 0.02(4) & 1.69(3) & 0.63(2) & 1.61(5) & & $\approx 1$ & $\approx 0$ \\
   exp. (flat) & 1.67(5) & 0.03(3) & 1.64(3) & 0.60(3) & 1.67(5) & & $\approx 1$ & $\approx 0$ \\
   Eden (circular) & 1.67(2) & 0.00(3) & 1.68(3) & 0.67(2) & 1.61(9) & 0.63(3) & 1.02(4) & 0.01(3) \\
   PNG (flat) & 1.672(15) & 0.010(15) & 1.664(13) & 0.667(13) & 1.64(2) & 0.64(4) & 1.00(2) & 0.001(12) \\ \hline
   theory (table~\ref{tbl:ExpKPZ}) & 5/3 & 0 & 5/3 & 2/3 & 5/3 & 2/3 & 1 & 0 \\ \hline
  \end{tabular}
  }\\
 \raggedright\scriptsize $^{\rm a}$ The number in parentheses indicates a range of error in the last digit(s), evaluated within the observation time. The symbol $\approx$ represents that the data are consistent with the indicated exponent value, but do not seem to reach the asymptotic time region during the observation time; consequently, the range of error could not be estimated reliably. \\
 \raggedright\scriptsize $^{\rm b}$ Abbreviations: exp. = liquid-crystal experiment, Eden = off-lattice Eden model, PNG = discrete PNG model.
 \end{center}
\end{table}

To evaluate the $T$-dependence of $S_q(\omega; T)$,
 we seek for such a value of $z_q$ that $S_q(\omega; T) T^{z_q}$ overlaps
 with different $T$ (\figref{fig5}).
While the experimental data seem to favor $z_q \approx 1/2$
 within the limited observation time (\figref{fig5}a-d),
 the numerical data, obtained with much wider ranges of $T$,
 overlap reasonably well with $z_q \approx 2/3$ for large $T$
 (\figref{fig5}e,f and insets;
 see also table~\ref{tbl:ExpRes} for the estimates)
 and rule out $z_q = 1/2$ (\figref{fig5}g,h and insets).
Indeed, since $q(x,t) \sim \mathcal{O}(1)$,
 the integrated power $\int_{2\pi/T}^\infty S_q(\omega;T) \rd\omega$
 should remain finite for this process
 \cite{Mandelbrot-book1999,Niemann.etal-PRL2013,Leibovich.etal-PRE2016};
 with $S_q(\omega;T) \sim T^{-z_q} \omega^{-\alpha_q}$,
 this condition implies $z_q = \alpha_q - 1 = 2/3$.

\begin{figure}[p]
 \centering
 \includegraphics[width=\hsize,clip]{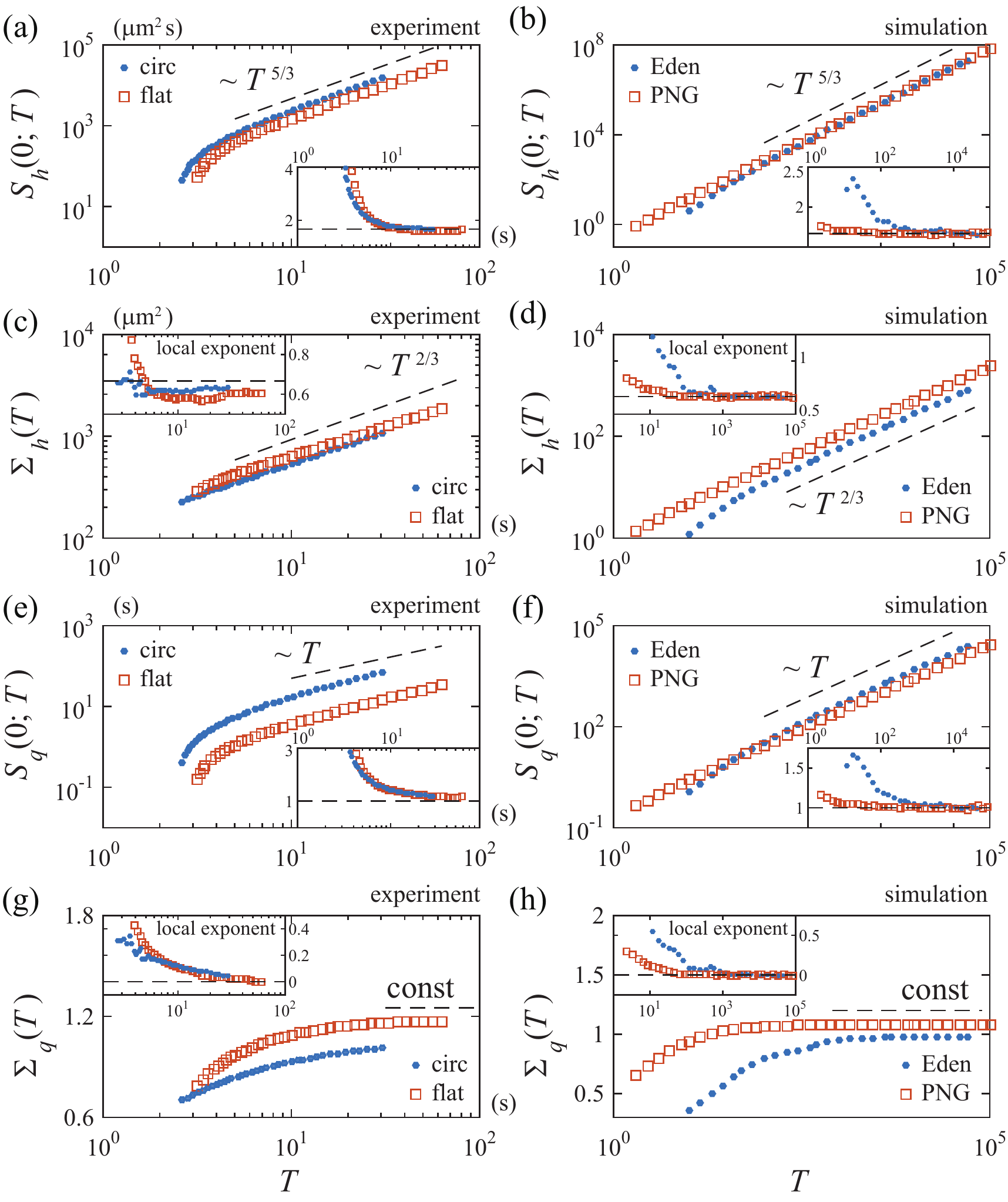}
 \caption{
Evaluation of the exponents $\mu_*$ and $\delta_*$ for the liquid-crystal experiments (a,c,e,g) and for the simulations (b,d,f,h). Data for the circular and flat interfaces are shown in blue dots and red squares, respectively. The dashed lines are guides for the eyes indicating the exponent expected from the aging Wiener-Khinchin theorem. The insets show the local exponent $\diff{(\log y)}{(\log T)}$ against $T$, where $y$ denotes the ordinates used in the main panels.
}
 \label{fig6}
\end{figure}

Further, recent studies \cite{Sadegh.etal-NJP2014,Leibovich.Barkai-PRL2015,Dechant.Lutz-PRL2015,Leibovich.etal-PRE2016} proposed
 a set of critical exponents for the power spectrum.
With $S_*(\omega; T)$ given by \eqref{eq:DefPowSpec}, 
 where the subscript $*$ represents the choice of the variable $X(t)$,
 the exponents $(\alpha_*, z_*, \mu_*, \delta_*)$ are defined by:
\begin{equation}
\expct{S_*(\omega;T)} \sim T^{-z_*} \omega^{-\alpha_*} \quad (\mathrm{for~small~} \omega)
 \label{eq:Exp1}
\end{equation}
 and
\begin{equation} 
 \expct{S_*(0;T)} \sim T^{\mu_*}, \qquad
 \Sigma_*(T) \equiv \int_{2\pi/T}^\infty \expct{S_*(\omega;T)} \rd\omega \sim T^{\delta_*}.
  \label{eq:Exp2}
\end{equation}
For our study, we define two sets of the exponents
 $(\alpha_h, z_h, \mu_h, \delta_h)$ and $(\alpha_q, z_q, \mu_q, \delta_q)$
 for $S_h(\omega;T)$ and $S_q(\omega;T)$, respectively.
The results in figures~\ref{fig4} and \ref{fig5} then indicate
 $\alpha_h = \alpha_q = 5/3$, $z_h = 0$, and $z_q = 2/3$.
For the exponents $\mu_*$ and $\delta_*$, our data indicate
 $\mu_h = 5/3$, $\delta_h = 2/3$, $\mu_q = 1$, and $\delta_q = 0$
 (\figref{fig6} and table~\ref{tbl:ExpRes}),
 though the experimental results for some exponents are not conclusive
 within the limit of the observation time.

\subsection{Aging Wiener-Khinchin theorem}

In fact, the exponents $(\alpha_*, z_*, \mu_*, \delta_*)$ we obtained
 can be accounted for
 by the recently formulated ``aging Wiener-Khinchin theorem''
 \cite{Leibovich.Barkai-PRL2015,Dechant.Lutz-PRL2015,Leibovich.etal-PRE2016}.
Similarly to the standard Wiener-Khinchin theorem
 valid for stationary processes, the aging Wiener-Khinchin theorem describes
 relationship between the power spectrum $S_*(\omega; T)$ and
 the correlation function $C_*(\tau; t)$,
 defined by \eqref{eq:DefPowSpec} and \pref{eq:DefCorrFunc}, respectively,
 for aging processes.
Specifically, under the assumption that the correlation function $C(\tau; t)$
 takes the form
\begin{equation}
 C_*(\tau; t) \simeq t^{\Upsilon_*} \phi_*(\tau/t)  \label{eq:CorrFuncRes}
\end{equation}
 for large $\tau$ and $t$,
 which is usually satisfied in scale-invariant systems
 including the KPZ-class interfaces,
 the aging Wiener-Khinchin theorem
 states that the ensemble-averaged power spectrum
 $\expct{S_*(\omega; T)}$ is obtained by
\begin{equation} \fl
 \expct{S_*(\omega; T)} = \frac{2T^{\Upsilon_*+1}}{2+\Upsilon_*} \int_0^1 \rd \zeta (1-\zeta)^{\Upsilon_*} \phi_*\(\frac{\zeta}{1-\zeta}\) {\textstyle {}_1F_2\[1\!+\!\frac{\Upsilon_*}{2}; \frac{1}{2}, 2\!+\!\frac{\Upsilon_*}{2}; -(\frac{\omega T\zeta}{2})^2\]}  \label{eq:AgingWK}
\end{equation}
 with the hypergeometric function ${}_1F_2[a; b_1, b_2; x]$
 \cite{Leibovich.Barkai-PRL2015,Dechant.Lutz-PRL2015,Leibovich.etal-PRE2016}.
Then one can show that the exponents $(\alpha_*, z_*, \mu_*, \delta_*)$
 are controlled by short-time behavior of $\phi_*(\tau/t)$,
 which can usually be expanded as
\begin{equation}
 \phi_*(y) \simeq a_* - b_* y^{V_*} \quad (y \ll 1),  \label{eq:DefPhi}
\end{equation}
 and obtain the following scaling relations \cite{Leibovich.etal-PRE2016}:
\begin{eqnarray}
 &\alpha_* = 1+V_*, \qquad
 &z_* = V_* - \Upsilon_*, \nonumber \\ 
 &\mu_* = 1 + \Upsilon_*, \qquad
 &\delta_* = \max\{ -z_*, \mu_*-1 \}.  \label{eq:AgingWK1}
\end{eqnarray}

For the KPZ class, with time series $X(t) = \delta h(x,t)$ or $q(x,t)$,
 we have \cite{Krug.etal-PRA1992}
\begin{equation}
 \Upsilon_h = 2\beta = 2/3, \quad
 \Upsilon_q = 0,  \quad
 V_h = V_q = 2\beta = 2/3,
\label{eq:ExpKPZ}
\end{equation}
 where we used $\beta = 1/3$ for the $(1+1)$-dimensional case.
With \eqref{eq:AgingWK1} and \pref{eq:ExpKPZ}, we obtain
 the values of $(\alpha_h, z_h, \mu_h, \delta_h)$
 and $(\alpha_q, z_q, \mu_q, \delta_q)$
 as summarized in table~\ref{tbl:ExpKPZ},
 which are consistent with all the observations presented
 in figures~\ref{fig4}-\ref{fig6} and table~\ref{tbl:ExpRes}.

\begin{table}[b]
 \begin{center}
  \caption{Critical exponents for the $(1+1)$-dimensional KPZ class.}
  \label{tbl:ExpKPZ}
  \catcode`?=\active \def?{\phantom{0}}
  {
  \begin{tabular}{l|cc|cccc} \hline
   time series & $\Upsilon_*$ & $V_*$ & $\alpha_*$ & $z_*$ & $\mu_*$ & $\delta_*$ \\ \hline
   height fluctuations $\delta h(x,t)$ & 2/3 & 2/3 & 5/3 &  0  & 5/3 & 2/3 \\
   rescaled height $q(x,t)$       &  0  & 2/3 & 5/3 & 2/3 &  1  &  0  \\ \hline
  \end{tabular}
  }
 \end{center}
\end{table}

\subsection{$1/f^\alpha$ spectrum and the Baik-Rains universal variance}

In addition to the exponents,
 the aging Wiener-Khinchin theorem \pref{eq:AgingWK}
 also gives the coefficient $C_*$ of the $1/f^\alpha$ spectrum
 \cite{Leibovich.Barkai-PRL2015,Dechant.Lutz-PRL2015,Leibovich.etal-PRE2016},
 $\expct{S(\omega; T)} \simeq C_* T^{-z_*} \omega^{-\alpha_*}$
 with $\omega T \gg 1$, by
\begin{equation}
 C_* = \frac{2b_* \sin(\pi V_*/2) \Gamma(1+V_*)}{1+\Upsilon_* - V_*},
 \label{eq:AgingWK2}
\end{equation}
 with the gamma function $\Gamma(\cdot)$.
The condition $\omega T \gg 1$ implies that
 the time scale of interest, $\omega^{-1}$, is much smaller than
 the observation time $T$, albeit larger than any microscopic time scale
 of the system.

Interestingly, for KPZ, this condition is exactly
 what one would need for the Brownian subclass
 \cite{Krug.etal-PRA1992,Takeuchi-PRL2013,Ferrari.Spohn-SIG2016,DeNardis.etal-PRL2017}.
Starting from an arbitrary initial condition, 
 we wait for long time $T$, so that the interface profile $h(x,T)$
 becomes sufficiently close to the asymptotic profile, or Brownian motion,
 $B_x$ (see section~\ref{sec:KPZ}).
Now we regard $h(x,T)$ as a new initial condition
 and consider the relative height $h'(x,t') = h(x,T+t') - h(x,T)$,
 then $h'(x,t')$ evolves as in \eqref{eq:Rescaling}
 and the corresponding random variable $\chi'_{x,t'}$
 exhibits the universal properties
 of the Brownian KPZ subclass, as long as $t' \ll T$
 \cite{Takeuchi-PRL2013,HalpinHealy.Lin-PRE2014,HalpinHealy.Takeuchi-JSP2015,Ferrari.Spohn-SIG2016,DeNardis.etal-PRL2017}.
Indeed, with $C_h(\tau;t) = \expct{\delta h(t+\tau) \delta h(t)}$
 and $\phi_h(\tau/t)$ defined by \eqref{eq:CorrFuncRes} accordingly,
 we can show, for $\tau/t \to 0$,
\begin{equation}
 \phi_h(\tau/t) \simeq \Gamma^{2/3}\[ \cum{\chi_i^2} - \frac{1}{2}\cum{\chi_0^2}(\tau/t)^{2/3} + \mathcal{O}(\tau/t) \].  \label{eq:KPZCorrFunc}
\end{equation}
Here, $\chi_i = \chi_2$ or $\chi_1$ depending on the initial condition
 (circular or flat, respectively) and $\cum{\cdot^2}$ denotes the variance;
 hence $\cum{\chi_0^2}$ is the variance of the Baik-Rains distribution,
 a universal hallmark of the Brownian KPZ subclass (table~\ref{tbl:subclass}).
Comparing \eqref{eq:KPZCorrFunc} with \eqref{eq:DefPhi}, we obtain
 $b_h = \frac{1}{2}\Gamma^{2/3}\cum{\chi_0^2}$.
Therefore, using \eqref{eq:AgingWK2} with $V_h = \Upsilon_h = 2/3$,
 we finally arrive at
\begin{equation}
 S_h(\omega;T) \simeq \cum{\chi_0^2} C_h' \omega^{-5/3}, \qquad
 C_h' = \frac{\sqrt{3}}{2}\Gamma(5/3) \Gamma^{2/3}.  \label{eq:AgingWK3}
\end{equation}
In other words, the coefficient of the $1/f^\alpha$ spectrum turns out to be
 the universal Baik-Rains variance, times a constant factor $C_h'$.

\begin{figure}[p]
 \centering
 \includegraphics[width=\hsize,clip]{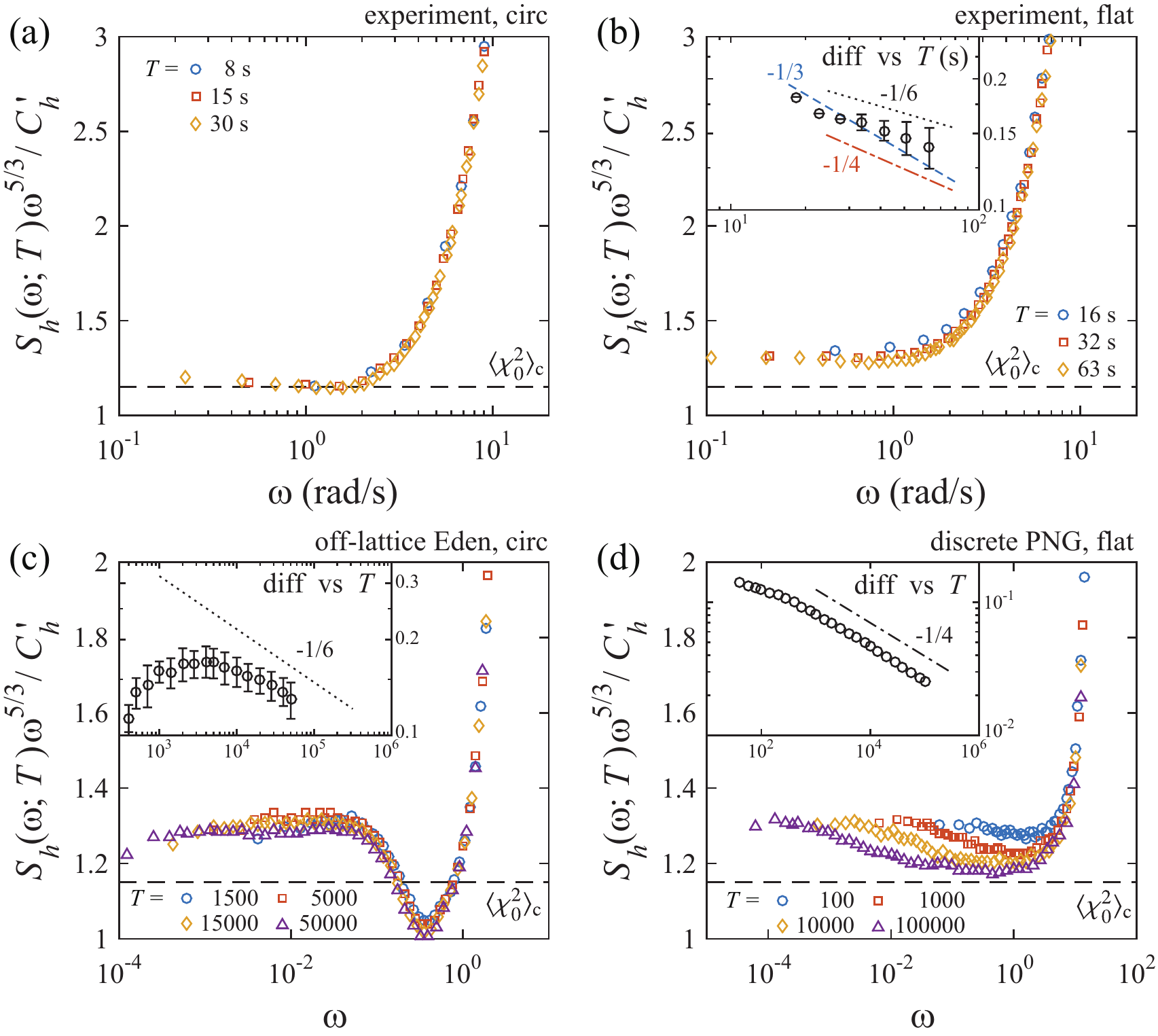}
 \caption{
Comparison of the coefficient of the $1/f^\alpha$ spectrum with the Baik-Rains universal variance, for the liquid-crystal experiments [circular (a), flat (b)], the off-lattice Eden model [circular (c)], and the discrete PNG model [flat (d)]. According to the aging Wiener-Khinchin theorem and the KPZ scaling laws, the ordinate $S_h(\omega;T)\omega^{5/3}/C_h'$ is expected to converge to the Baik-Rains variance $\cum{\chi_0^2}$ for $T^{-1} \ll \omega \ll \min\{\Delta t^{-1},\tau_0^{-1}\}$, where $\Delta t$ is the time resolution and $\tau_0$ is a microscopic time scale of the system (see \eqref{eq:AgingWK3}). The dashed lines indicate $\cum{\chi_0^2} = 1.15039$. In the insets, the difference between $S_h(\omega;T)\omega^{5/3}/C_h'$ and $\cum{\chi_0^2}$ is plotted as a function of $T$. The difference is measured from the value of the plateau region of the main-panel data for (b,c), with errorbars indicating the minimum and maximum in that region; for (d) the difference is measured from the local minimum estimated by cubic fitting. The dotted, dashed-dotted, and dashed lines are guides for the eyes, indicating exponents -1/6, -1/4, and -1/3, respectively. 
}
 \label{fig7}
\end{figure}

This is tested in \figref{fig7},
 where $S_h(\omega;T)\omega^{5/3}/C_h'$ is plotted against $\omega$
 and compared with $\cum{\chi_0^2}$.
For the experimental circular interfaces, we find remarkable agreement
 (\figref{fig7}a); from the local minimum of the plateau,
 we obtain $1.151(12)$, very close to the Baik-Rains variance
 $\cum{\chi_0^2} = 1.15039$ \cite{Prahofer.Spohn-PRL2000}.
For the other three cases (\figref{fig7}b-d), in contrast,
 we find some deviations from $\cum{\chi_0^2}$,
 which however decrease with increasing $T$.
The deviations are found to decay by some power law $T^{-a}$,
 with exponent $a$ compatible in the range $1/6 \lesssim a \lesssim 1/3$.
From the PNG data, $a \approx 1/4$ is suggested, but
 data for other systems and for longer observation times are needed
 to make a conclusion about the value of this exponent.
In any case, all the results presented in \figref{fig7} indeed
 support the validity of \eqref{eq:AgingWK3} in the asymptotic limit;
the coefficient of the $1/f^\alpha$ spectrum is essentially
 the Baik-Rains universal variance.

Our result indicates that,
 if the relative height $h'(x,t') = h(x,T+t') - h(x,T)$ with $t' \ll T$
 is considered,
 circular and flat KPZ interfaces indeed approach the Brownian subclass.
This crossover has been known for the flat interfaces
 \cite{Takeuchi-PRL2013,HalpinHealy.Lin-PRE2014,HalpinHealy.Takeuchi-JSP2015,Ferrari.Spohn-SIG2016},
 but to the knowledge of the author it is first shown here
 for the circular case,
 apart from indirect evidence in \cite{DeNardis.etal-PRL2017},
 in agreement with theoretical predictions
 \cite{Ferrari.Spohn-SIG2016,DeNardis.etal-PRL2017}.
Curiously, by direct analysis of the relative height $h'(x,t')$,
 using the liquid-crystal experimental data for the flat case,
 the variance of the rescaled relative height
 remained far from the Baik-Rains variance within the observation
 time \cite{Takeuchi-PRL2013},
 which is also the time used in the present study.
As shown here, for this experimental system,
 the power spectrum exhibits the Baik-Rains signature much earlier,
 by simpler analysis.
On the other hand, the earlier work \cite{Takeuchi-PRL2013} also
 used a discrete PNG model%
\footnote{
The discrete PNG model studied in \cite{Takeuchi-PRL2013} was
 defined with the synchronous update, while in the present paper
 we used the checkerboard update. The author also measured
 the power spectrum $S_h(\omega;T)$ in the case of the synchronous update,
 and found a result similar to \figref{fig7}.
},
 for which the Baik-Rains variance was more easily found
 in the relative height $h'(x,t')$ than in the power spectrum
 shown in the present paper.
Indeed, the relative height method has been usually used
 in numerical studies and provided good estimates
 of the Baik-Rains variance \cite{Krug.etal-PRA1992,Tang-JSP1992,Takeuchi-PRL2013,HalpinHealy.Lin-PRE2014,HalpinHealy.Takeuchi-JSP2015}.
Since the two methods seem to have different strengths
 of the finite-time effect, 
 it would be useful to measure
 both the relative height and the power spectrum.
Note that the power spectrum method can also be used for higher dimensions,
 for which the counterpart of the Baik-Rains distribution
 is not solved analytically.
For $2+1$ dimensions, numerical \cite{HalpinHealy-PRE2013}
 and non-exact theoretical \cite{Kloss.etal-PRE2012} estimates are available;
 it would be then interesting to compare the power-spectrum method
 with these known results.

\section{Concluding remarks}  \label{sec:conclusion}

In this paper, we have studied the $1/f^\alpha$-type power spectrum
 of interface fluctuations in the $(1+1)$-dimensional KPZ class,
 both experimentally and numerically,
 using the liquid-crystal turbulence as well as simple models,
 specifically, the off-lattice Eden model and the discrete PNG model.
We measured the power spectrum for two different sets of time series,
 namely those for the height fluctuation $\delta h(x,t)$
 and the rescaled height $q(x,t)$,
 and characterized it by a set of critical exponents
 (figures~\ref{fig4}-\ref{fig6} and table~\ref{tbl:ExpRes}).
The observed exponents were found to be in good agreement with
 predictions \pref{eq:AgingWK1} and \pref{eq:AgingWK2}
 from the recently proposed, aging Wiener-Khinchin theorem
 \cite{Leibovich.Barkai-PRL2015,Dechant.Lutz-PRL2015,Leibovich.etal-PRE2016}.
Moreover, this theory revealed that
 the coefficient of the $1/f^\alpha$ spectrum contains essential information
 on the universal properties of the asymptotic KPZ interfaces,
 namely the Baik-Rains universal variance, through \eqref{eq:AgingWK3}.
This intriguing connection was corroborated
 by both experimental and numerical data (\figref{fig7}).
In relation to the KPZ universality subclass (table~\ref{tbl:subclass}),
 our results constitute experimental and numerical supports that
 the relative height $h'(x,t') = h(x,T+t') - h(x,T)$
 of circular and flat interfaces
 belongs to the Brownian KPZ subclass in the asymptotic limit $t'/T \to 0$
 with $t', T \to\infty$.
On the other hand, the power spectrum does not seem to be
 as efficient for studying time correlation properties
 in the other limit $t'/T \to \infty$,
 which characterize the subclass for the bare height $h(x,t)$
 (i.e, circular or flat in our case)
 and are known to be different between the two cases, even qualitatively
 \cite{Takeuchi.Sano-JSP2012,Ferrari.Spohn-SIG2016,Takeuchi.Akimoto-JSP2016}.

From broader perspectives, it may be worth making a few general remarks 
 about the power spectrum and the stationarity,
 which are not new but sometimes overlooked in the literature.
First, one should recall that
 the stationarity in the power spectrum ($z_* = 0$) does not
 necessarily imply the stationarity of the underlying process,
 neither the boundedness of the time series;
 this is well-known from examples like the Brownian motion
 and reminded here in \figref{fig4}a-d,
 where $S_h(\omega;T)$ does not depent on $T$
 but the process $\delta h(x,t)$ is aging and unbounded.
The stationarity of the process can be inferred instead
 by measuring in addition the ``power'' at zero frequency,
 $\expct{S(0;T)} = T\expct{\bar{X}^2}$
 with $\bar{X} = \frac{1}{T}\int_0^T X(t) \rd t$,
 which is either linear in $T$ (if $\bar{X} \neq 0$)
 or constant (if $\bar{X} = 0$) for stationary processes,
 while in the aging case it grows with a nontrivial exponent
 $\mu_* = \alpha_* - z_*$.
Consistency with the aging scenario can be checked further
 through the set of the scaling relations \pref{eq:AgingWK1}
 from the aging Wiener-Khinchin theorem
 \cite{Leibovich.Barkai-PRL2015,Dechant.Lutz-PRL2015,Leibovich.etal-PRE2016}.
It is also important to remark that, in aging systems,
 time average and ensemble average are in general different \cite{Niemann.etal-PRL2013,Leibovich.Barkai-PRL2015,Dechant.Lutz-PRL2015,Leibovich.etal-PRE2016};
 while the present work focused on the ensemble average,
 a different form of the aging Wiener-Khinchin theorem should be used
 if one aims to connect to the time-averaged correlation function.
Since aging seems to occur rather often
 in systems showing $1/f^\alpha$ noise,
 neither the equivalence between ensemble average and time average
 nor the usual Wiener-Khinchin theorem should be used
 unless firm evidence of stationarity is obtained.
From this perspective,
 measuring the exponents $(\alpha_*, z_*, \mu_*, \delta_*)$
 is a versatile approach
 that does not require any assumption on aging or stationarity,
 yet useful to characterize underlying processes,
 as shown for the example of the KPZ class we studied here.
It is hoped that this approach will also help to study
 other examples of the $1/f^\alpha$-type spectrum,
 reported in the wealth of physical systems
 \cite{Dutta.Horn-RMP1981,Wong-MR2003,Musha.etal-book1992,Mandelbrot-book1999}.

\ack

The author wishes to thank E. Barkai, A. Dechant, and T. Halpin-Healy
 for their enlightening discussions,
 as well as critical reading of the manuscript
 by R. A. L. Almeida, E. Barkai, A. Dechant, and T. Halpin-Healy.
The author acknowledges financial support
 by KAKENHI from JSPS, Nos. JP25103004, JP16H04033, JP16K13846.

\section*{References}
\bibliographystyle{iopart-num}
\bibliography{agingWK}

\end{document}